# Method for critical current angular dependencies analysis of superconducting tapes

V V Guryev, S V Shavkin and V S Kruglov

National Research Centre "Kurchatov institute", 123182, Kurchatov sq. 1, Moscow, Russia

E-mail: GuryevVV@mail.ru

**Abstract.** Various applications of superconducting materials require accounting of anisotropy of the current-carrying capacity relative to magnetic field direction - $I_c(\theta)$. However, today there is no sufficiently comprehensive model that takes into account the anisotropy, therefore the angular dependences are usually not analysed, but only described using various mathematical formulas. As a result, the fitting parameters have no physical meaning and it is difficult to correlate the picture with the features of the microstructure. In this paper, we propose a method for analysing the critical current angular dependences based on the anisotropic pinning model. The applicability of this model for conventional superconducting Nb-Ti tapes with one peak in the Ic ($\theta$) dependence is shown. The possibility of extending this model to analyse the angular dependences of HTS materials is discussed.

## 1. Introduction

The pronounced anisotropy of the current-carrying capacity in relation to the magnetic field direction, inherent in promising high-temperature superconducting tapes (so called HTS coated conductors), arouses a great interest in methods for the critical current angular dependencies analysis. The angular dependence is understood as the critical current dependence taken in the geometry of the experiment in which the transport current and the magnetic field are perpendicular to each other (i.e. in maximum Lorentz force configuration), with a change in the angle $\theta$ between the magnetic field and the normal to the superconducting tape. The dependence obtained in this way is used, on the one hand, as a part of the material equation in the design of superconducting devices; on the other hand, it can provide valuable information about the pinning landscape and serves to develop the technology of the superconductor fabrication themselves.

It should be noted that the crystallographic unit cell of HTS ($ReB_2C_3O_{7-x}$ or ReBCO, where Re means rare earth element) is perovskite-like layered orthorhombic with similar *a*- and *b*- parameters and a larger ~3 times *c*- parameter; therefore, it is significantly anisotropic. This leads to the anisotropy of the effective masses of charge carriers in the normal state $\gamma = \sqrt{m_c/m_{a,b}} \sim 5-7$. It was shown that the upper critical field as thermodynamic parameter is anisotropic with respect to the magnetic field direction and controlled by $\gamma$ through the functional dependence $f(\theta) \propto (\cos^2\theta + \gamma^{-2}\sin^2\theta)^{-1/2}$ [1, 2]. Note that the critical current density is a kinetic parameter and determined not by thermodynamics but by the pinning landscape. Nevertheless, in the particular case of a weak collective pinning by point pins which is typical for HTS single crystals, the critical current density also has an angular dependence similar to $f(\theta)$ [3]. The superconducting layer of the coated conductor has a biaxial texture; therefore, in the case of weak pinning, the abovementioned tendencies may take place. However, with improved pinning, one can expect a significant deviation from $f(\theta)$, which was indeed often observed experimentally [4-10]. Thus, the functional dependence $f(\theta)$ (which is often used to analyze the critical current angular dependences) in case of practical coated conductors with high performance has no theoretical justification, nor is supported by experimental observations and so it is not legitimate.

In addition, several descriptive functional dependencies with a number of fitting parameters that have no physical meaning have been proposed [11-17]. If several peaks appear on the angular dependence, then the lonely peaks are considered first, and then their combination with the best fit of the weight coefficients is used [18,19]. In such approach, there is still little room for the analysis, since the correct description of the peaks shape can carry some statistical information about the distribution of the pinning centers [20-23]. Apparently, the best description of the experimental data can be achieved by completely entrusting the fitting procedure to the computational algorithms [24]. However, in the latter case, analysis is almost impossible, since there is no functional dependence as such.

In this paper we describe a method to analyze the angular dependence based on the previously developed anisotropic pinning model (APM) [25-27]. In the second section the theoretical background and the derivation of the formula used to analyze the critical current angular dependences are given. In the third section the experimental data obtained on the Nb-Ti tape are presented and analyzed. This tape compare favorably with coated conductors in that it has a cubic unit cell, so it does not have "internal" anisotropy associated with the effective mass, and all anisotropy is "external" due to the pinning landscape. The pinning center system for these tapes is well known and therefore it is the practically ideal model material. Then, in the final section, we present our thoughts on how this analysis can be extended to the case of coated conductors.

## 2. Theoretical background

The APM model considers an ensemble of a sufficiently large number of vortices as something whole, located in the so-called cooperative potential well due to the entire ensemble of pinning center. Let's assume that the vortices ensemble occupies the most favorable place at the bottom of the potential well. This condition *does not* imply that each of the vortices captured by individual pinning centers and stands at the bottom of its own individual potential well. Under the influence of the transport current, the vortices are jostled along the well slope, and if the Lorentz force is greater than the maximum slope angle of the well, the vortices start to move with constant velocity. The maximum pinning force, which counteracts the Lorentz force, is defined as:

$$\mathbf{F}_p = -\max(\frac{\partial U}{\partial \mathbf{l}}) = -\mathbf{e}_l \frac{U_0(\mathbf{B})}{L_0(\mathbf{j}, B)} \qquad (1)$$

where $U$ is the cooperative potential well depth, $\mathbf{e}_l$ is the unit vector in the direction of the Lorentz force, $L_0(\mathbf{j}, B) = U_0(\mathbf{B})/|\max(\partial U/\partial \mathbf{l})|$ is the effective size of the cooperative potential well, $U_0(\mathbf{B})$ - the effective depth of the cooperative potential well.

Thus APM works in the critical state approach and the real smearing of current-voltage characteristic is ignored. An essential assumption of the model is that the effective depth of the cooperative potential well $U_0(\mathbf{B})$ is assumed to depend only on the magnitude and direction of the magnetic induction (and does not depend on the current density and, therefore, the Lorentz force), and the size of the cooperative potential well $L_0(\mathbf{j}, B)$ is assumed to depend on the absolute value of the magnetic induction (density of the vortices) and the direction of the Lorentz force. The effective size of the cooperative potential well $L_0$ is determined by the minimum distance between two energetically equivalent positions of the vortex matter. As an estimate of this distance, it can be taken the minimum of two values: the average inter-vortex distance $a = \sqrt{4\Phi_0/\sqrt{3}B}$, where $B$ and $\Phi_0$ – induction and magnetic flux quantum, or the typical distance between pins in the direction of vortex motion.

According to the model all pinning features in a particular material are determined by the specific angular and field dependences of the depth $U_0$ and the size $L_0$. It was shown [25] that in the case of cold-rolled Nb-Ti tape, the extreme values of $U_0$ and $L_0$ are achieved with orientations of magnetic induction and Lorentz force along the main orthogonal directions in the material: normal direction (ND), rolling direction (RD) and direction perpendicular to rolling in the plane of the tape (PD). For other directions, $U_0$ and $L_0$ smoothly vary from minimum to maximum. Thus, the depth $U_0$ and the width $L_0$ of the cooperative potential well are described by ellipsoids:

$$\left(\frac{\cos\alpha''}{U^{RD}}\right)^2 + \left(\frac{\cos\beta''}{U^{PD}}\right)^2 + \left(\frac{\cos\gamma''}{U^{ND}}\right)^2 = \frac{1}{U_0^2} \quad (2a)$$

$$\left(\frac{\cos\alpha'}{L^{RD}}\right)^2 + \left(\frac{\cos\beta'}{L^{PD}}\right)^2 + \left(\frac{\cos\gamma'}{L^{ND}}\right)^2 = \frac{1}{L_0^2} \quad (2b)$$

where cos $\alpha''$, cos $\beta''$ and cos $\gamma''$ are the direction cosines of induction vector $\boldsymbol{B}$; cos $\alpha'$, cos $\beta'$ and cos $\gamma'$ are the direction cosines of the Lorentz force vector $\boldsymbol{F_L} = [\boldsymbol{J} \times \boldsymbol{B}]$; $U^{RD}$, $U^{PD}$, $U^{ND}$, $L^{RD}$, $L^{PD}$ and $L^{ND}$ – depth and width of the cooperative potential well in the main orthogonal directions. Since at the experiment only the ratio of the model parameters is measured, it is impossible to determine their absolute values from transport measurements.

If the consideration of the problem is reduced to the angular dependence geometry described in previous section, we obtain the following expression for the pinning force:

$$F_p(\theta) = j_c(\theta)B = \frac{U_0(\theta)}{L_0(\theta)} = \frac{U^{PD}}{L^{ND}}\sqrt{\frac{\left(\frac{L^{ND}}{L^{PD}}\cos\theta\right)^2+(\sin\theta)^2}{\left(\frac{U^{PD}}{U^{ND}}\cos\theta\right)^2+(\sin\theta)^2}} \quad (3)$$

or in terms of critical current density:

$$j_c(\theta) = j_c(90°)\sqrt{\frac{(k^L\cos\theta)^2+(\sin\theta)^2}{(k^U\cos\theta)^2+(\sin\theta)^2}} \quad (4)$$

where $k^L = \frac{L^{ND}}{L^{PD}}$ and $k^U = \frac{U^{PD}}{U^{ND}}$ – can be considered as the dimensionless parameters of fitting the critical current angular dependence. From the physical meaning of the coefficients, it follows that their ratio sets the ratio of critical currents when the field is oriented in the plane ($\theta = 90°$) and along the normal ($\theta = 0°$): $k^U/k^L = j_c(90°)/j_c(0°)$. Figure 1 shows examples of angular dependencies for the same $k^U/k^L$ ratio. The functional dependences of isotropic $U_0$ case (1/0.2) differ significantly from the isotropic $L_0$ case (5/1), and thus the study of the angular dependence makes it possible to separate the contributions of these quantities.

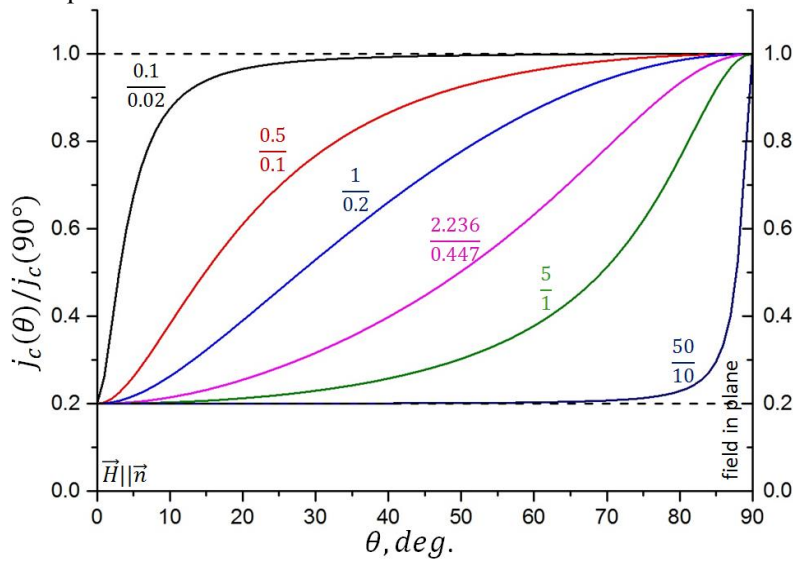

**Figure 1.** Functional angular dependence (4) for various parameters $k^U$ and $k^L$ with a constant $k^U/k^L$ ratio.

## 3. Results and discussions

### 3.1. Experiment details

In this paper we presents the results obtained on a superconducting Nb-Ti tape, whose structure we have studied in detail earlier by both electron microscopy [28, 29] and X-ray [30] methods. It was found that, as a result of cold rolling, the NbTi grains are strongly elongated in the rolling direction and oblate in the normal direction. The grain size distribution obeys an asymmetric log normal distribution with the most probable sizes: in the normal direction – 38 nm, in the direction perpendicular to rolling – 0.17 μm. It is well known that vortex pinning in single-phase Nb-Ti superconductors occurs at grain boundaries.

The Nb-Ti tape is 10 microns thick and has 1 micron copper stabilization on each side. To study the angular dependence, we cut out 2-mm-wide samples along the rolling direction. The measurements were carried out in liquid helium environment by the transport method on a special insert probe which allowed sample rotation. The magnetic field was provided by a superconducting solenoid up to 13 T. The critical current $I_c$ was determined using the "standard criterion" for an electric field of 1 μV/cm. It was previously shown [25,31] that all the effects associated with the inhomogeneous distribution of the current over the cross section of the sample due to the self-field end in fields above 0.2 T, therefore, the critical current density was calculated using a simple relation $j_c = I_c/S$, where $S$ is the sample cross-sectional area. The pinning force was defined as $F_p = j_c\mu_0 H$.

### 3.2. Angle and field dependencies

The inset to Fig. 2 shows the critical current angle dependences of the Nb-Ti tape for three fields: 0.7 T, 5 T, and 9 T. For convenience of comparison, in the main part of Figure 2, these dependencies are rebuilt in relative units. The dotted lines correspond to the fitting curve (4) with the fitting parameters from Table 1.

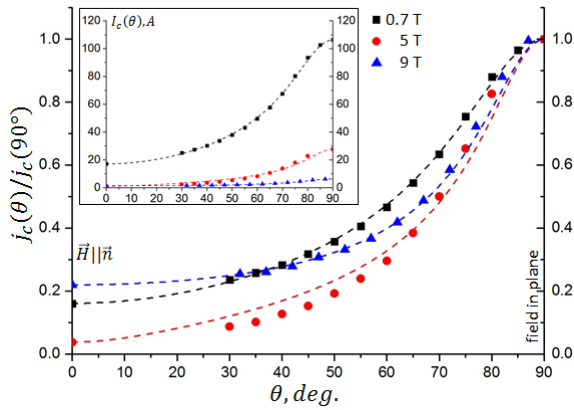
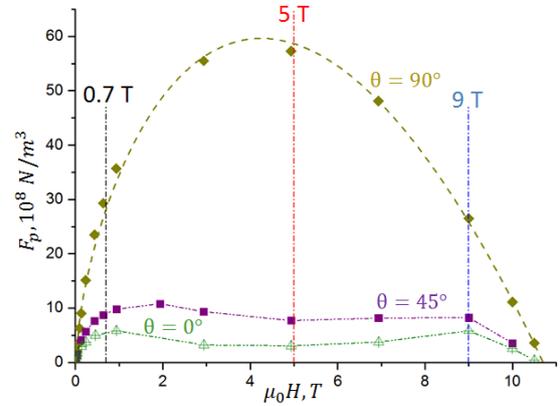

**Figure 2.** Experimental angular dependences of the reduced critical current density for the Nb-Ti tape in 0.7 T, 5 T and 9 T (dots), and model angle dependences (4) (dashed lines). The inset shows the angular dependence of the critical current in absolute units.

**Figure 3.** Pinning field dependencies for three different magnetic field orientations: $\theta = 0°$ (field normal to the tape), $\theta = 45°$, $\theta = 90°$ (field in plane of the tape).

The huge difference in the degree of anisotropy for fields of 0.7 and 9 T compared to 5 T is noteworthy. If for the former the ratio $k^U/k^L$ is equal to 6.2 and 4.6, respectively, then for the latter the degree of anisotropy $k^U/k^L$ reaches 26.5 (Table 1). This is due to crossover to weak pinning in perpendicular geometry [27]. Indeed, when the field is in the tape plane, the pinning field dependence has a domed shape, which is typical for strong pinning (Fig. 3). In contrast to this, in the perpendicular geometry the pinning field dependence has a two-humped form corresponding to crossovers from strong to weak pinning and back [32]. The volume pinning force demonstrates maximum values in perpendicular field of about 1 T and 9 T [27]. These points can be taken as estimates for crossover boundaries. Thus within angular dependence measured at 5 T (Fig. 2) there is a transition area from weak ($\theta = 0°$) to strong pinning ($\theta = 90°$). Another feature of this angular dependence, probably related to the previous one is the quite high standard deviation (Table 1). In extreme positions, substantially different physics of vortex matter pinning is realized; therefore, it is rash to expect a good description of such complex processes from the simple model. For 5 T, the anisotropy of the depth $U_0$ and width $L_0$ of the cooperative potential well make an almost equal contribution to the angular dependence of the critical current.

**Table 1.** Fitting parameters of the model, standard deviation and vortex distance.

|  | 0.7 T | 5 T | 9 T |
|---|---|---|---|
| $k^U$ | 3.42 | 5.03 | 4.78 |
| $k^L$ | 0.55 | 0.19 | 1.05 |
| Standard deviation | $8.0 \times 10^{-3}$ | $42.9 \times 10^{-3}$ | $9.5 \times 10^{-3}$ |
| $a$, nm | 83 | 26 | 19 |

For 0.7 T and 9 T, the contribution of $U_0$ significantly exceeds the contribution of $L_0$. For 9 T, the case of almost isotropic width $L_0$ is realized. This suggests that the typical value of the distance between the pinning centers is larger than inter vortex spacing for any possible direction of the vortices motion. This conclusion is supported by microscopic data, since the minimum typical size of Nb-Ti grains (~38 nm) is larger than the mean inter-vortex distance at 9 T (~19 nm). At the same time, the anisotropy of the potential well $U_0$ is in good agreement with the anisotropy of the typical grain shape $d^{PD}/d^{ND}$=4.5 which indicates an individual and strong pinning.

In contrast at 0.7 T the vortex matter is rare enough and easy adapts to pinning centers, and the strong pinning mode is realized for all orientations of the magnetic field. For this field value, the vortex distance (~68 nm) is greater than the grain size in the normal direction, but less than the size in the direction perpendicular to rolling (~170 nm). This leads to anisotropy not only in the depth $U_0$, but also in the width of the potential well $L_0$. In this case, we note that the anisotropy $k^L$ determined as a fitting parameter to the angular dependence is in good agreement with the assumptions of the model, namely, it corresponds to the ratio of the vortex spacing and the grain size in the normal direction.

## 4. Further development and concluding remarks

In this paper we present the method for analyzing of the critical current angular dependence of superconducting tapes based on the simple anisotropic pinning model. We have shown that this analysis is applicable quite good to a cold-rolled Nb-Ti tape with a well-known system of pinning centers. Next we discuss the issues to extend this analysis to HTS coated conductors. The fundamental assumption of the abovementioned model is the fact that the width of the cooperative potential well $L_0$ is independent of the magnetic induction direction, which may not be valid in materials with internal anisotropy of superconducting properties such as in HTS. In HTS single crystal one can distinguish two different cases regarding flux lattice configuration when the Lorentz force $\boldsymbol{F_L} = [\boldsymbol{J} \times \boldsymbol{B}]$ acts in the same direction (for example along $b$-axes): 1) if the magnetic field directed along $c$-axes (perpendicular to basic plane) and transport current directed along $a$- axes; and 2) if the magnetic field directed along $a$-axes (in basic plane) and transport current directed along $c$-axes. For the first case, the dimension of the vortex core is determined by the coherence length $\xi_{a,b}$, in the second one - $\sqrt{\xi_c \xi_{a,b}}$. So the mean distance between «effective» pinning centers in these cases may be different. This hypothetically leads to a change in $L_0$- parameter of the model.

Another issue is the shape of the $U_0$ and $L_0$ bodies in case of HTS coated conductors in which specific artificial pinning centers (APC) like nanorods or radiation tracks often occurred. However, direct shape correction based on the often observed multi-peak angular dependencies [33-35] may lead to incorrect results, since the observation of some peaks depends on the electric field threshold [36, 37]. The "true" value of latter may be less [38] or more [39] than the conventionally chosen 1 µV/cm.

Hence, in order to restore the current-carrying capacity of a superconductor in an arbitrary magnetic field, it is necessary not only to know the evolution of three-dimensional bodies of $U_0$ and $L_0$, but also to implement an upgraded anisotropic pinning model taking into account the shape of the current-voltage characteristic.